# Legal and Ethical Implications of Applications based on Agreement Technologies: The Case of Auction-based Road Intersections

**José-Antonio Santos**[*], **Alberto Fernández**[+], **Mar Moreno Rebato**[*],
**Holger Billhardt**[+], **José A. Rodríguez García**[*], **Sascha Ossowski**[+]

Contact author: José-Antonio Santos (joseantonio.santos@urjc.es)

**Abstract** Agreement Technologies refer to a novel paradigm for the construction of distributed intelligent systems, where autonomous software agents negotiate to reach agreements on behalf of their human users. Smart Cities are a key application domain for Agreement Technologies. While several proofs of concept and prototypes exist, such systems are still far from ready for being deployed in the real-world. In this paper we focus on a novel method for managing elements of smart road infrastructures of the future, namely the case of auction-based road intersections. We show that, even though the key technological elements for such methods are already available, there are multiple non-technical issues that need to be tackled before they can be applied in practice. For this purpose, we analyse legal and ethical implications of auction-based road intersections in the context of international regulations and from the standpoint of the Spanish legislation. From this exercise, we extract a set of required modifications, of both technical and legal nature, which need to be addressed so as to pave the way for the potential real-world deployment of such systems in a future that may not be too far away.

## 1. Introduction

The transactions and interactions among people in modern societies are increasingly mediated by computers. From email, over social networks, to virtual worlds, the way people work and enjoy their free time is changing dramatically. The resulting networks are usually large in scale, involving huge numbers of interactions, and are open for the interacting entities to join or leave at will. People are often supported by software components of different complexity to which some of the corresponding tasks can be delegated. In practice, such systems cannot be built and managed based on rigid, centralised client-server architectures, but call for more flexible and decentralised means of interaction. Accordingly, Brynjolfsson and McAfee (2016, p 96) say: "We can't predict exactly what new insights, products, and solutions will arrive in the coming years, but we are fully confident that they'll be impressive. The second machine age will be characterized by countless instances of machine intelligence and billions of interconnected brains working together to better understand and improve our world. It will make mockery out of all that came before." A new era that is in its infancy and that moves at a vertiginous pace towards an uncertain future.

In this context, machines with ability to show intelligent behaviour have turned what was strange to humans into an artificial naturalness. That is, these increasingly autonomous systems are becoming part of our daily routine and life is not understood without them. As Billhardt et al. (2015) argue, Artificial Intelligence (AI) challenges human work and tasks will be increasingly carried out in close collaborations between humans and machines creating so called human-agent teams.

From the technical point of view, the emerging field of Agreement Technologies (AT) (Ossowski et al., 2013) aims at supporting this vision. It aims at next-generation open distributed systems, where

---

[*] Faculty of Law and Social Sciences, University Rey Juan Carlos, Madrid, Spain

[+] CETINIA, University Rey Juan Carlos, Móstoles (Madrid), Spain

[*] Faculty of Law and Social Sciences, University Rey Juan Carlos, Madrid, Spain

[+] CETINIA, University Rey Juan Carlos, Móstoles (Madrid), Spain

[*] Faculty of Law and Social Sciences, University Rey Juan Carlos, Madrid, Spain

[+] CETINIA, University Rey Juan Carlos, Móstoles (Madrid), Spain

interactions between software components are based on the concept of agreement, and which enact two key mechanisms: a means to specify the *space* of agreements that the agents can possibly reach, and an interaction model by means of which agreements can be effectively reached. Autonomy, interaction, mobility and openness are key characteristics that are tackled from a theoretical and practical perspective.

There is a broad variety of domains where the potential of AT becomes apparent (see Part VII of Ossowski et al. 2013). In these domains, the choices and actions of a large number of autonomous stakeholders need to be coordinated, and interactions can be regulated, by some sort of intelligent computing infrastructure (Omicini et al., 2004), through some sort of institutions and institutional agents (Fornara et al., 2013), or simply by strategically providing information in an environment with a significant level of uncertainty (Centeno et al., 2009). The advent of intelligent road infrastructures, with support for vehicle-to-vehicle and vehicle-to-infrastructure communications, make smart transportation a challenging field of application for AT, as it allows for a decentralised coordination of individually rational commuters. The work by Vasirani and Ossowski (2012) on networks of auction-based road intersections, which account for preferential use of road infrastructures depending on drivers' profiles and preferences, is an excellent example for this.

In this context, interaction between human beings and machines can enable an individualised and at the same time fair access to public resources, which will increase the efficiency of such smart infrastructures and thus has the potential to improve both individual and social welfare. Still, such a scenario can be considered as yet another case of a cultural revolution fuelled by technological progress, with potentially ambivalent consequences. On the one hand, it shows a wide range of possibilities for action and possibilities of choice that can help improve the level of welfare of drivers. For example, in *semi-autonomous driving systems*, the passenger would be legally authorised to be distracted by phone calls or use social networks. Autonomous vehicles may also allow for carrying goods in less time by maintaining a more constant speed and without the obligatory breaks required for drivers. On the other hand, all sorts of driving-related jobs (taxi, bus, lorry, etc) would be lost. In addition, *decreased traffic accidents* would probably mean fewer insurance claims and fewer repairs to the garages. In any case, it is important to notice that, even though the technology is (almost) ready for those visions to become reality, there are numerous ethical and legal issues that need to be addressed before experimentation with those systems in the real world will become an option. In this matter, it is necessary to remember that no decision made in the field of law and ethics is trivial: decisions and their consequences need to be analysed within the force field among law, technology and ethics.

As Casanovas (2013) outlines, especially novel applications based on AT have a wide range of legal question and problems, which must be addressed if such systems are ever meant to be applied to the real-world. In this paper we focus on AT applied to smart intersection management, so as to illustrate this claim through a case study. In particular, setting out from the auction-based intersection managers introduced by Vasirani and Ossowski (2012), we identify and analyse legal and ethical problems related to the potential real-world deployment of this AT-based technology. Furthermore, we design and discuss solutions to overcome these problems from both perspectives: what modifications would be needed in the smart traffic infrastructure itself, and what changes in legal regulations would be required, if we were to deploy auction-based intersections in our society in a properly regulated manner.

This approach is related but intrinsically different to the inverse process of creating systems or applications that comply with legal requirements and which is usually treated in the field of legal requirements engineering (e.g. Boella et al., 2014; Rabinia and S. Ghanavati, 2018). In legal requirements engineering research the focus is usually on specifying languages and methods to specify legal regulations, in order to identify their implications on a new system. Here, we want to specify those aspects of a given system that may need modifications or adaptations in the existing regulations. We think this is a common problem, which is often not treated in a systematic manner. Especially in academic research, new solution approaches are usually proposed without analysing their viability from legal or ethical points of view. Thus, a posterior analysis in this direction should be or has to be carried out before transferring such approaches to commercial systems.

Our paper is structured as follows: Section 2 describes the technological foundations of an envisioned future road traffic infrastructure, where intersections are managed by smart controllers while drivers can reserve space/time slots at intersections according to their needs, and highlights those aspects of the approach that could affect or could be affected by legal issues. Section 3 analyses legal and ethical implications of such auction-based road intersections in the context of international regulations and from the standpoint of the Spanish legislation. From this exercise, in Section 4 we identify aspects from the proposal presented in Section 2 that do not comply with the current regulations and discuss possible solutions, of both technical and ethical/legal nature, which would need to be implemented in order to pave the way for the potential real-world deployment of such systems in a future that may not be too far away. Section 5 summarises our findings and points to future work.

## 2. Coordination of traffic flows through intelligent intersections

Removing the human driver from the control loop through the use of autonomous vehicles integrated with an intelligent road infrastructure can be considered as the ultimate, long-term goal of the set of systems and technologies grouped under the name of Intelligent Transportation Systems (ITS). The advantages of such an integration span from improved road safety to a more efficient operational use of the transportation network. For instance, vehicles can exchange critical safety information with the infrastructure, so as to recognise high-risk situations in advance and therefore to alert drivers. Furthermore, traffic signal systems can communicate signal phase and timing information to vehicles to enhance the use of the transportation network.

In this regard, many authors have recently paid attention to the potential of a tighter integration of autonomous vehicles with the infrastructure for intersection management (Namazi et al. 2019). Dresner and Stone (2008) introduce the *reservation-based* control system, where an intersection is regulated by an intelligent software component, called *intersection manager*, which assigns reservations of space and time to each autonomous vehicle intending to cross the intersection. Each vehicle is operated by another software agent, called *driver* agent. When a vehicle approaches an intersection, the driver agent asks the intersection manager to reserve the necessary space-time slots to safely cross the intersection. The latter simulates the vehicle´s trajectory inside the intersection and informs the driver agent whether its request is in conflict with the already confirmed reservations. If such a conflict does not exist, the driver agent stores the reservation details and tries to meet them; otherwise it may try again at a later time. The authors show through simulations that in situations of balanced traffic, if all vehicles are autonomous, their delays at the intersection are drastically reduced compared to traditional traffic lights.

Vasirani and Ossowski (2012) extend the above model for reservation-based intersection control along two major lines. Firstly, for *single intersections*, they elaborate an auction-based policy for the allocation of reservations to vehicles that grants preferential access to vehicles based on the drivers´ different attitudes regarding their travel times. Secondly, for *network of intersections*, they set up a computational market where intersection managers compete with each other as suppliers of reservations, and selfishly adapt the reserve prices of the auctions that they run, so as to match the actual demand. In this manner, traffic flows are better distributed among the road network, and the social cost of applying a preferential assignment strategy at intersection-level is reduced. In the following we summarise the dynamics and functionalities of this approach, as it sets the ground for the ethical and legal questions analysed in the rest of this article.

### 2.1. Mechanism for single intersection

For a single reservation-based intersection, the problem that an intersection manager has to solve comes down to allocating reservations among a set of drivers in such a way that a specific objective is maximised. This objective can be, for instance, minimising the average delay caused by the presence of the regulated intersection. In this case, the simplest policy to adopt is allocating a reservation to the first agent that requests it, as occurs with the first-come first-served (FCFS) policy proposed by Dresner and Stone in their original work. Another work in line with this objective takes inspiration from adversarial queuing theory for the definition of several alternative control policies that aim at minimising the average delay (Vasirani and Ossowski, 2009).

However, these policies ignore the fact that in the real world, depending on the context and their personal situation, people value the importance of travel times and delays quite differently. Since processing the incoming requests to grant the associated reservations can be considered as a process of assigning resources to agents that request them, intersection managers are supposed to allocate the disputed resources to the agents that value them the most. In line with findings from mechanism design, it is assumed that the more a human driver is willing to pay for the desired set of space-time slots, the more they value the good. Thus, the policy for the allocation of resources relies on *combinatorial auctions*.

The goods assigned through these combinatorial auctions are the right to use certain space inside the intersection at a given time. An intersection is modelled as a discrete matrix of space slots. Let *S* be the set of the intersection space slots, and *T* the set of future time-steps, then the set of items that a bidder can bid for is $I = S \times T$. Due to the nature of the problem, a bidder is only interested in *bundles* of items over the set *I*. As Figure 1 illustrates, in the absence of acceleration in the intersection, a reservation request implicitly defines which space slots at which time the driver needs in order to pass through the intersection. A bid over a bundle of items is implicitly defined by the reservation request. Given the parameters *arrival time*, *arrival speed*, *lane* and *type of turn*, the auctioneer (i.e., the intersection manager) is able to determine which space slots are needed at which time. In addition, the *value of its bid*, i.e. the amount of money that the driver is willing to pay for the requested reservation, must be included into a vehicle's reservation request.

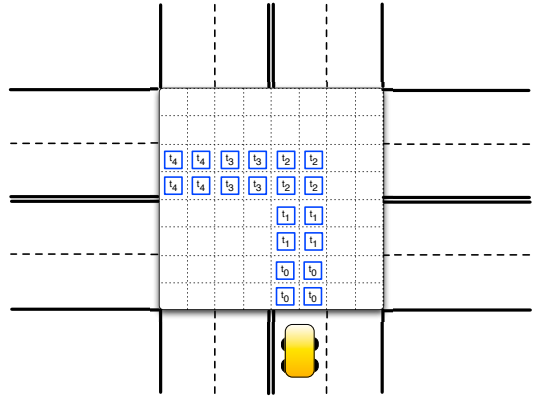

Figure 1. Bundle of items for a reservation request
(from Vasirani and Ossowski, 2012)

Figure 2 shows the interaction protocol used to regulate the combinatorial auction. It starts with the auctioneer waiting for bids for a certain amount of time. Once the new bids are collected, they constitute the bid set. Then, the auctioneer executes an anytime approximation algorithm to solve the winner determination problem (WDP), thus determining the winner set with the bids whose reservation requests have been accepted. During the WDP algorithm execution, the auctioneer still accepts incoming bids, but they will only be included in the bid set of the next round. The auctioneer sends a *CONFIRMATION* message to all bidders that submitted the bids contained in the winner set, while a *REJECTION* message is sent to the bidders that submitted the remaining bids. Then a new round begins, and the auctioneer collects new incoming bids for a certain amount of time.[1]

---

[1] The protocol also allows vehicles to cancel a reservation that they already have acquired for a certain time *t*. This may happen, for instance, when a driver realises that, due to changing traffic conditions, it is likely to arrive at the intersection at some time *t'>t*. The auction protocol includes additional constraints that keep agents from strategizing based on this option.

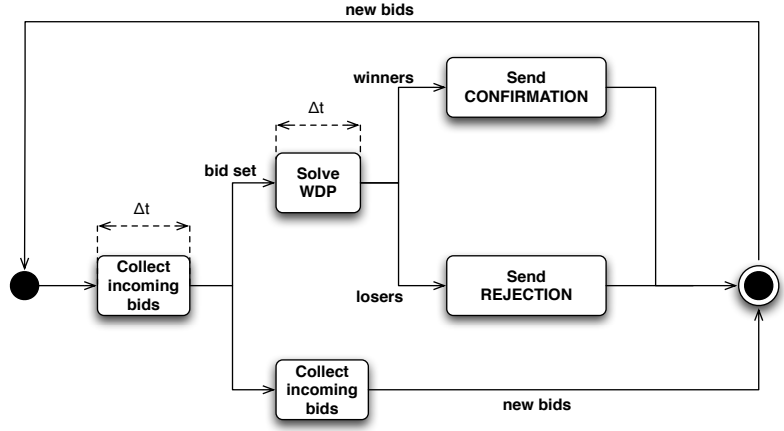

**Figure 2.** Auction policy proposed by Vasirani and Ossowski (2012)

Through a set of simulated experiments, Vasirani and Ossowski (2012) confirmed that the policy based on combinatorial auctions (CA) enforces an inverse relation between the amounts spent by the bidders and their *delay* (the increase in travel time due to the presence of the intersection). That is, the more money a driver is willing to spend for crossing the intersection, the faster will be its transit through it. They noticed that even with a theoretically infinite amount of money, a driver cannot experience zero delay when approaching an intersection, as the travel time is influenced by slower potentially *poorer* vehicles in front of it. They also confirmed that the CA policy has a higher *average* delay compared to FCFS strategy, since it grants a reservation to the driver that values it the most, rather than maximising the number of granted requests. This *social cost* of CA is the bigger, the higher the load of the intersection, i.e. the more drivers compete for reservations. The extension of the CA mechanism to multiple intersections described in the sequel aims at reducing this *social cost* of giving preference to drivers with a high valuation of time.

## 2.2. Mechanisms for multiple intersections

In case of a single intersection, a driver agent simply needs to decide on the preferred value for the bid that it submits to the auctioneer. The decision space of a driver agent in an urban road network with *multiple* intersections is much broader: complex and mutually dependent decisions must be taken such as route choice and departure time selection. Therefore, this scenario opens up new possibilities for intersection managers to affect the behaviour of drivers. For example, an intersection manager may be interested in influencing the collective route choice performed by the drivers, using variable message signs, information broadcast, or individual route guidance systems, so as to evenly distribute the traffic over the network. This problem is called *traffic assignment*.

In the *Competitive Traffic Assignment / Combinatorial Auction* strategy (CTA-CA), each intersection manager runs the CAs described in the previous section, but includes a reserve price, i.e. it announces a minimum price that it charges for the reservations that it sells. At each time $t$, there may be different reserve prices $p^t(l)$ for each of the incoming link $l$ of the intersection. The reserve prices of the intersections are made available to driver agents in real time, which then choose their routes according to their personal preferences about travel times as well as the current monetary costs.

Each intersection manager competes with all others for the supply of the reservations that are traded. It applies a simple reserve price update rule that aims at attracting drivers to the intersection when traffic demand is low, and at deterring them from choosing routes that transit through the intersection when it is oversaturated. At time $t+1$, the reserve price on link $l$ is computed as follows:

$$p^{t+1}(l) = p^t(l) + p^t(l)\frac{z^t(l)}{s(l)}$$

In this formula, the constant $s(l)$ represents the optimal amount of vehicles that can cross the intersection coming from link $l$. The excess demand $z^t(l)$ expresses the difference between total demand at time $t$ and supply $s(l)$ on link $l$ (notice that $z^t(l)$ becomes negative when the supply on link $l$ exceeds the demand). By dynamically adjusting the reserve prices of the network's intersection managers, traffic flows are better distributed among the network, thus avoiding an over-saturation of *bottleneck* intersections.

Vasirani and Ossowski (2012) evaluated the approach using simulations upon a topology that resembled the high capacity road network of the city of Madrid in Spain. Among others, they compared CTA-CA with a FCFS strategy, where each intersection manager assigns reservations in isolation on a first-come first-served basis. Notice that differently from CTA-CA, on the one hand, in FCFS there is no notion of *social cost* as all drivers are treated equally, but on the other hand FCFS cannot benefit from the traffic assignment effects of the reserve prices of CTA-CA. In the experiments, with the FCFS strategy, the drivers´ route choice model simply selects the route with minimum expected travel time at free flow, since there is no notion of price. For the evaluation of the CTA-CA strategy, it is assumed that drivers choose the most preferred route they can afford. The experiments measure the moving average of the travel time, that is, how the average travel time of the entire population of drivers, computed over all the origin-destination pairs, evolves during the simulation.

The results show that CTA-CA outperforms FCFS, while maintaining the inverse relation between the amounts spent by driver agents and their vehicle's *delay*. The performance gain of CTA-CA is the bigger, the higher the load at the intersections. These results indicate that preferential assignment methods such as CTA-CA can provide an individualised access to the public traffic infrastructure that is tailored to a driver's preferences and needs, while their negative effect on social welfare (i.e. increased *average* travel time) can be mitigated by adequately designing the assignment mechanisms.

### 2.3. Description of system operation

While the previous sections summarised the technical feasibility of auction-based intersections and studied their physical performance from both individual and social perspectives, we now aim at analysing the compliance of the proposed intersection management system with current legal regulations. The objective is to identify necessary and possible modifications, not only in the legal regulations but also in the smart infrastructure itself, that would allow to properly implement the system in society.

In order to allow for a legal analysis of a technical solution (as presented in this paper) essentially three steps have to be carried out:

1. *System operation specification*: Identification and specification of all aspects of the proposed solution that could affect or could be affected by legal issues. On the one hand, such a specification should include the high-level application domain(s) of the system as well as all affected entities. And on the other hand, it should specify the principal operations (that may affect legal regulations) but hiding away technical details. The aim is to allow legal experts to analyse compliance with current regulations.
2. *Compliance analysis*: Legal experts should analyse whether the different aspects identified in the previous step match current regulations. This step is often not straightforward and may require a more detailed analysis of different interpretations of existing norms. If a violation of existing norms is detected, the violated norms or legal principles should be identified and associated to the corresponding aspect of the system.
3. *Modification proposals*: Based on the analysis and discussion in step 2, possible solutions should be identified that would make the proposed system compliant. Such solutions may be proposals of modifications of the existing regulations, modifications that have to be introduced in the proposed system, or hybrid solutions. Here, the effects or consequences of modifying each part (regulation or system) have to be pondered so as to decide which one should be adapted. For example, in the case of a detected incompliance with fundamental rights adapting the regulation may be unfeasible, thus the system modification would be the only option to make it compliant.

Even though formal languages with rigorous semantics and proof theory could be used to specify the operation principles of the proposed system, for the aforementioned purpose in step 1 this is not really necessary. We rather require that the specification should be "understandable" by legal experts in order to allow for the analysis and the discussions in steps 2 and 3, so a structured natural language description seems more appropriate. Upon this background, in the sequel we summarise key aspects of the operation of autonomous intersections and networks presented in sections 2.1 and 2.2 as follows:

- Domain: Traffic.
- Subdomain: regulation of paths of vehicles through intersections.
- Affected entities: vehicles, intersection manager.
- Definitions:
  - Intersection Manager: virtual entity that regulates the paths of vehicles through a specific intersection. Different intersections are regulated by different intersection managers.
  - Vehicles: vehicles that want to pass through an intersection.
  - Space-time slot: a space slot that is part of a route through an intersection from one incoming road to an outgoing road.
- Operational facts:
  - Fact 1: A vehicle can only pass through an intersection using the space-time slots assigned to it previously by the intersection manager and using the assigned space slots at the specified time.
  - Fact 2: The intersection manager assigns at most one vehicle to each space slot at each particular time.
  - Fact 3: In order to claim a path through an intersection, any vehicle has to request space-time slots from the corresponding intersection manager.
  - Fact 4: If more than one vehicle requests the same space-time slots, the assignment is based on an auction between the vehicles. The type of auction employed is one-shot, sealed bid. The winner of an auction pays the bid to the intersection manager and the intersection manager assigns the corresponding space-time slots to the winner.
  - Fact 5: The intersection manager makes benefit.
  - Fact 6: Each intersection has a reserve price which is published by the intersection manager. In order to participate in an auction, each vehicle's bid has to be higher than the current reserve price of the intersection.
  - Fact 7: The intersection manager aims at improving the distribution of the traffic by specifying the reserve price. The intersection manager establishes this price locally, incrementing it when the demand grows. There is no maximum price.
  - Fact 8: Each intersection manager takes its decisions locally, without explicit coordination with other managers in the neighbourhood. Thus, in scenarios with high general traffic density in a particular area and limited number of alternative paths, reserve prices at all intersections in that area may grow.
- Assumptions:
  - Assumption 1: All vehicles that want to cross an intersection need to be capable to communicate with the intersection manager and have to have the software system installed.
- Affected general issues:
  - Responsibility (in case of failure, or illicit behaviour).
  - Data protection.

## 3. Legal issues and concerns

The proposal described in the preceding pages of this article regarding smart intersections that grant reservations of space and time to drivers based on electronic online auctions gives rise to numerous legal-administrative problems within the framework of the current European and Spanish legislation, which stems from the Vienna Convention on Road Signs and Signals (1968). In the following we will carry out a *legal science fiction* exercise to point out some of the many legislative changes that would need to be introduced into these legal systems to enable the deployment of intelligent road infrastructure of such characteristics[2], beginning with the modification of traffic regulations (Unger-Sternberg, 2018).

### 3.1. Auction-based intersections and administrative law in Spain

The technology is prepared, or almost ready, to deploy auction-based intersections, but the law not yet. Still, we know that the law always goes one step behind the reality that it aims to regulate. The United States is the pioneer, in terms of legislation, regarding the incipient regulation of autonomous vehicles and intelligent road infrastructures.[3] For examples, just consider the state laws of Nevada,[4] California,[5] or Michigan.[6]

European Union Law defines *Intelligent Transport Systems* (ITS) as "advanced applications which without embodying intelligence as such aim to provide innovative services relating to different modes of transport and traffic management and enable various users to be better informed and make safer, more coordinated and 'smarter' use of transport networks." The European Union identifies the following priority areas for the development and use of ITS:[7]

- Optimal use of road, traffic and travel data,
- Continuity of traffic and freight management ITS services,
- ITS road safety and security applications,
- Linking the vehicle with the transport infrastructure.

In Spain, an intelligent road infrastructure that uses auction-based intersections would be subject, in the first place, to the road legislation. In order for this road infrastructure to be installed and used, such as the smart intersection based on auctions, in Madrid (Spain), as stated in this paper (section 2), it is necessary to know that Spain is a politically decentralised country where political power resides in three territorial levels: The Spanish State, the Autonomous Communities and the local governments. In Spain, these three territorial spheres have competence on roads. Consequently, this autonomous intersection can be used on a highway whose owner is the State, the Autonomous Community of Madrid or the municipality of Madrid. In the proposal of a single intersection (section 2.1) that collects this paper it is necessary to determine to which public administration the road belongs because each territorial administration has elaborated its own regulations on roads, hence the importance of knowing the existing territorial division in Spain.

The State Highway Law promotes research and development of intelligent infrastructures of this type. In this sense, article 1 of the Spanish Law 37/2015, of September 29, of Roads has the following purposes:

"b) Offer the necessary infrastructure for the transport of people or goods.

d) To obtain an offer of road infrastructure and services associated to them of quality, safe and efficient, with an adequate allocation of resources.

e) Promote research, development and technological innovation, as well as its dissemination.

g) Encourage the development of advanced services to mobility and road transport."

---

[2] The Directorate-General of Traffic of Spain has signed a cooperation agreement with the company Mobileye to start testing autonomous vehicles on Spanish roads, https://newsroom.intel.com/news-releases/mobileye-spains-road-safety-authority-dgt-collaborate-enhance-road-safety/. Vision Zero technology of this company is important when developing autonomously managed intersections especially those where bicycles and pedestrians circulate, ... vid. Towards Vision Zero: Intelligent Intersection Infrastructure to enhance safe operations of (self-driving) cars, https://its.berkeley.edu/node/13121.

[3] *National Conference of States Legislatures*, http://www.ncsl.org/research/transportation/autonomous-vehicles-self-driving-vehicles-enacted-legislation.aspx and *Self-Drive Act,* Sec. 5, https://www.congress.gov/bill/115th-congress/house-bill/3388/text

[4] Nevada Revised Statutes § 482A, https://www.leg.state.nv.us/NRS/NRS-482A.html

[5] California Vehicle Code § 38750, https://leginfo.legislature.ca.gov/faces/codes_displaySection.xhtml?lawCode=VEH§ionNum=38750

[6] Michigan Vehicle Code § 257.2.b.; http://www.legislature.mi.gov/(S(z4ly25pwxohcwibffe5szntb))/mileg.aspx?page=GetObject&objectname=mcl-257-2b

[7] Directive 2010/40/EU of the European Parliament and of the Council, of 7 July 2010, on the framework for the deployment of Intelligent Transport Systems in the field of road transport and for interfaces with other modes of transport. In Spain, Royal Decree 662/2012, of April 13, on the framework for the deployment of Intelligent Transport Systems in the field of road transport and for interfaces with other modes of transport.

In any case, intelligent road infrastructures would need important incentives and investments, for example, the current Smart Cities and Islands Plan of Spain should take these investments into account and prioritise them.[8]

However, also in this area, not only the competences on roads matter, but other State competences are important as well, such as:
- "public works of general interest or whose implementation affects more than one Autonomous Community" (Judgement of the Spanish Constitutional Court No. 65/1998);
- the general communications regime;
- the road system that passes beyond the country borders;

In matters of "traffic and circulation of motor vehicles" (article 149.1.21 of the Spanish Constitution), "are not encompassed only by the conditions related to traffic (for example, traffic signals, speed limitations, etc.), but the conditions to be met by vehicles that circulate by roads", because the scope of this subsection of article 149.1.21a reaches any guarantees of safety in circulation (Judgments of the Spanish Constitutional Court No. 59/1985 and, subsequently, No. 181/1992). Therefore, when projecting this precept on the road matter, the State is also competent to establish all those rules of a technical nature that directly affect safety in traffic. This solution is supported, as indicated in the aforementioned Judgments, "in the fact that the guarantees of safety in circulation, according to the will of the Constitution ..., must be uniform throughout the national territory" and, consequently, the Spanish State must comply with international treaties on these matters, especially international conventions on road traffic. A vehicle can only pass through an intersection using the space-time slots assigned to it previously by the intersection manager and using the assigned space slots at the specified time. Traffic regulations, Spanish and international, should be modified to contemplate this possibility (facts 1 and 2).

The road legislation, both state and autonomous communities, considers that roads are ways of public domain and of public use, built and marked mainly for the circulation of motor vehicles (article 2.2 of the State Road Law and article 3 of the Law of Madrid of roads). Therefore, the proposal made in this paper (section 2.3 and fact 1) should bear in mind that roads and their intersections or crossroads are public goods properties, with all the legal consequences and limitations that this implies for an eventual smart intersection based on auctions, where drivers need to pay to reserve their time and space for transiting safely through it. Although, in principle, the proposed intersection would not contradict the consideration as public domain property of the roads, except for some points that will be discussed in this paper (fact 1).

For the legal-administrative regulation of these auction-based intersections we must rely on: the regulation on public goods properties, on the possible reversal of the public domain,[9] the possible financing of their construction (public, private, types of contracts for their tendering, ...), public management (direct or through a company or public entity); private management (concessionaire, payment of a fee, ...).[10] Also, a public law entity could be created and it would be subject to private law, a State Agency or a special company for the construction and exploitation of this intelligent infrastructure and, in particular, the autonomously managed intersections, which guarantees cooperation between the different public administrations, a fact that does not guarantee the proposal included in section 2.2. of this paper, multiple intersections, Fact 8. Finally, we could consider the complete privatization of the road network, although the legal doctrine is unanimous in considering that the common use, whose guarantee corresponds to the public authorities, would be very seriously compromised with the delivery of the road network to the private sector (Bobes Sánchez, 2007, pp 127-128).

---

[8] On Spanish Smart Cities Plan, 2015, vid. http://www.minetad.gob.es/turismo/es-ES/Novedades/Documents/Plan_Nacional_de_Ciudades_Inteligentes.pdf.; Estrategia española de i+d+i en inteligencia artificial, 2019, vid. http://www.ciencia.gob.es/stfls/MICINN/Ciencia/Ficheros/Estrategia_Inteligencia_Artificial_IDI.pdf. On AENOR Norms in relation to smart cities, vid. http://www.agendadigital.gob.es/planes-actuaciones/Bibliotecaciudadesinteligentes/Material%20complementario/normas_ciudades_inteligentes.pdf. Also, there are projects financed by the European Union in relation to intelligent road infrastructure, vid. the Inframix R & D initiative (Road Infrastructure ready for mixed vehicle traffic flows) https://cordis.europa.eu/project/rcn/210131_en.html and Digital Europe programme for the period 2021-2027, https://eur-lex.europa.eu/legal-content/EN/TXT/HTML/?uri=CELEX:52018PC0434&from=EN. For its part, the New York State Law encourages the establishment of communication technology that allows wireless-enabled infrastructure to share information electronically with vehicles and thereby facilitate the progress of the implementation of autonomous vehicles, vid. An Act to emend the transportation law, in relation to establishing a pilot program for vehicle-to-infrastructure technology, of the State of New York, January 30, 2018.

[9] Public domain goods cannot be privately owned; they are outside the private legal trade. They are public property. The decision to declare a category of goods as public goods properties belongs to the legislator through the Law (article 132 of the Spanish Constitution: 1. The legal system governing public domain and community property shall be regulated by law, on the principle that they shall be inalienable and imprescriptible and not subject to attachment or encumbrance. 2. The property of the State public domain shall be established by law and shall, in any case, include coastal area, beaches, territorial waters and natural resources of the economic zone and the continental shelf). Some examples of public goods properties, in Spain, are roads, surface and underground waters, ports, coasts, … (coasts because the Spanish Constitution says it). The reversal would suppose that the good is outside the public domain. The reversal of roads of the public domain would need a Law of reversal. It is a decision of the legislator.

[10] *Automated vehicles. Do we know which road to take?* Infrastructure Partnerships Australia, 2017.

The goods that the legislator configures as a public domain are susceptible to three types of uses: 1. The common or general use that any citizen can make, provided that others are not excluded, it is free and does not require any formal legal base. 2. The use especially intense, dangerous, profitable, or preferably in cases of scarcity that requires an administrative authorization or a concession contract, if its use exceeds four years and, 3. The private use that implies occupation of a portion of the public domain, excludes others and requires the granting of an administrative concession contract (articles 84, 85 and 86 of the Spanish Law 33/2003, of November 3, on the Patrimony of Public Administrations).

In this regard, it has been said that "the common use of the public domain corresponds to the exercise of public freedom. Thus the freedom to circulate on public roads is a manifestation of freedom of movement" (Bobes Sánchez, 2007, p 125). In other words, roads are part of the public domain (article 5 of the Spanish Law 33/2003, of November 3, the Patrimony of Public Administrations) and the circulation of vehicles is the common use of roads (Santamaría Pastor, 1985, p 390) which is the manifestation of freedom of movement in a motorised world "and it corresponds equally and indiscriminately to all citizens, so that the use by some does not impede the other interested ones," as says article 85 of the State Spanish Law on the Patrimony of Public Administrations. It would also be possible, as we said earlier, the special use and even the exclusive use of roads, as public domain goods (at least in some of their segments that include alternative routes).

In relation to the regulation of roads (including intersections), the principle of equality, in the use of the public domain, in our case the auction-based intersection, should be included its future regulation (Fact 1). However, both case-law and legal doctrine in relation to this principle of equality in the use of the public domain of roads have placed special emphasis on distinguishing identical factual assumptions that can support a legitimate difference in the use of the corresponding public domain as, for example, we can distinguish and, therefore, limit their use to neighbours and residents of certain roads of other users; also, we can differentiate depending on the tonnage of vehicles to prohibit the use of roads in certain hours and days. Other reasons of public interest are: the reduction of pollution or the regulation of traffic density to facilitate circulation, also allow introducing variables that can unequivocally suggest some type of discrimination, whose differentiation has a reasonable and objective justification (Bobes Sánchez, 2007, p 126). These issues related to the legal classification of roads and their intersections, as well as smart intersections such as the one included in this paper, as public domain goods, must guarantee the freedom of movement and the principle of equality. This does not preclude a special use that distinguishes, for example, among residents, among vehicles according to their tonnage, restricts or even prohibits the circulation of certain vehicles due to their degree of pollution or which can only circulate through roads fully autonomous vehicles to facilitate the circulation or avoid accidents. However, all these forecasts are not included in the algorithm of section 2.3 of this paper.

Does this intersection guarantee freedom of movement and equality? Does it allow a vehicle to cross the intersection without paying because there is no alternative route? In principle, the algorithm described in section 2 does not allow any vehicle to cross the intersection for free. This road infrastructure of autonomous intersection based on auctions, proposed in section 2, would violate the freedom of movement because the proposed algorithm does not guarantee that every vehicle without a driver can cross the intersection for free (fact 6). Therefore, the algorithm should be modified so that the user is informed of the waiting time to be able to pass for free. In this way, the user will assess their individual welfare, as explained in section 2: "the more money a driver is willing to spend to cross the intersection, the faster the transit will be through it". If the user is not willing to pay, the waiting time will be longer but the pass through the intersection will be not prohibited. Another option would be the construction or existence of an alternative route, for free, to the autonomous intersection.

Among these restrictions on the common use of the road public domain the regulations on traffic and road safety are included. This traffic regulation protects the public interest that involves the development of traffic in conditions of safety and fluency. This autonomous infrastructure, as indicated in this paper, has as main objectives to guarantee road safety and traffic flow. The proposed algorithm prioritises the traffic flow through the payment of a price determined by the auction (fact 7). The Spanish legislation on traffic has its main rule in the Royal Legislative Decree 6/2015, of October 30, which approves the revised text of the Law on Traffic, Circulation of Motor Vehicles and Road Safety. The current traffic regulations and road safety do not contemplate an infrastructure like the proposal of this paper and do not allow the use of roads by fully autonomous vehicles (assumption 1) (On Dutch traffic law, Prakken, 2017).[11]

Finally, unlike other proposals (for example, Dresner and Stone, 2008), the smart intersections proposed in this paper require payments for space / time reservations at the intersection, which would be adjusted through an auction mechanism. The type of auction proposed in this paper fits with the electronic auction, described in article 143 of the Spanish Law 9/2017, of November 8, on Contracts of the Public Sector (fact 4).

The proposal of autonomous intersection with payment for its use or space / time reservation implies obtaining benefits (fact 5). Although we have already indicated that the construction and management of this type of infrastructure can be carried out in different ways (public management / private management) we now focus on the issue of road tolls. The assumption would be as follows: the public Administration in charge of a highway awards the construction and operation of auction-based intersections to a private company that would be the concessionaire. The concessionaire, in exchange for building the infrastructure, acquires the exploitation of the intelligent infrastructure as compensation and imposes the payment of a road toll;[12] that is, the payment of the users for the use of this infrastructure. Although the public administration is competent to determine the maximum price, the algorithm proposed in section 2 must have a maximum limit because currently it does not contemplate it (fact 7). The payment could be justified by making sure that the common use of roads was not compromised, the fundamental freedom of movement, in this case that the only possibility of the exercise of freedom of movement was through toll roads. The algorithm must contemplate that it can be crossed free of charge and that there is a maximum price to cross the intersection (facts 6 and 7). The existence of the road toll could violate the principle of equality but not the "free of charge" principle, which, as such, does not integrate the concept of public domain (Bobes Sánchez, 2007, p 128).

The tolls for the use of road infrastructures, as the Spanish Supreme Court has indicated, are payments that users pay to the concessionaire of a service, they are not fees or fiscal obligations or payments for services rendered to the public within the meaning of article 31. 3 of the Spanish Constitution ("Personal or property contributions for public purposes may only be imposed in accordance with the law") but they are trade-offs for the service provided by the concessionaire that the concessionaire makes its own by private law; that is, through the private management of this road infrastructure benefits are obtained, without prejudice to the intervention that the granting Administration may have in setting it up in exercise of the corresponding the tariff regulation, different from taxing powers (Judgment of the Court Spanish Supreme Court of April 30, 2001, among others). The public Administration (state, autonomous or local) must establish a maximum price to pass through the intersection and, it can demand that the crossing be for free, in certain conditions like the nonexistence of alternative route (facts 6 and 7). It is not contrary to article 14 of the Spanish Constitution (principle of equality) [Spaniards are equal before the law and may not in any way be discriminated against on account of birth, race, sex, religion, opinion or any other personal or social condition or circumstance.], according to the Spanish Supreme Court, the difference in the payment of tolls at the time that some must satisfy it while other users are not required to do so to the same extent (Judgement of the Spanish Supreme Court of April 29, 2002, for example). Following this case-law of the Spanish Supreme Court[13], the design of the auction algorithms for crossing the smart intersection should include differentials in the payment of the road toll, which may have a surcharge, or a bonus, or even the total exemption depending on the type of vehicle. In this regard, Vasirani and Ossowski (2011, p 214) argue that the computational market could be enriched by allowing "intersection managers to offer a range of different pricing instruments (e.g., off-peak pricing, daily subscriptions, and fares based on usage of the intersection)". However, the proposal contained in section 2 does not do it. For example, total exemption vehicles might include traffic police, government police and other law enforcement and judicial authorities, as well as ambulances and fire services when they had to carry out some mission on the highway[14]... But the algorithm described in section 2 does not differentiate between types of vehicles, whether they are emergency or not. The algorithm must allow the crossing of these emergency vehicles first, before any other vehicle, and free of charge. Following the example of the Massachusetts, Bill S. no. 1945, 2017 (which allows the use of roads of that State of the United States by autonomous vehicles and a fee is established for their use).[15] Also, the existence of discounts in the payment of these road tolls could be considered: depending on the number of passengers carried by the autonomous vehicle, encouraging public transport or collaborative transport (door to door); depending on transit geographic areas where there are few public transport options available; based on income or personal income, establishing a higher rate depending on the tonnage of the heavy vehicle, etc. The proposed algorithm would increase the price in case of traffic congestion to infinity; that is to say, as we have indicated, it does not contemplate a maximum price that the Spanish public administrations should set. Finally, the proposed algorithm does not include any of these promotion measures for vehicles with high occupancy or less polluting; based on family income; etc.

---

[11] The non-law Proposition of the Spanish Popular Party states: "The Congress of Deputies urges the Government to establish an adequate legal framework that allows: ...
- Promote the development and use of the autonomous vehicle by developing specific legislation and classifying the possible legal gap posed by the introduction of the autonomous vehicle in circulation", (Official Gazette of the Spanish Parliament, No. 204, September 8, 2017, pp 32- 33).

[12] Directive 2004/52/EU of the European Parliament and of the Council, of 29 April 2004, on the interoperability of electronic road toll systems in the Community. This Directive lays down the conditions necessary to ensure the interoperability of electronic road toll systems in the Community. It applies to the electronic collection of all types of road fees, on the entire Community road network, urban and interurban, motorways, major and minor roads, and various structures such as tunnels, bridges and ferries. Systems of electronic toll collection which are put in place in the Member States should meet the following fundamental criteria: the system should be amenable to ready incorporation of future technological and systems improvements and developments without costly redundancy of older models and methods, the costs of its adoption by commercial and private road users should be insignificant compared with the benefits to those road users as well as to society as a whole, and its implementation in any Member State should be non-discriminatory in all respects between domestic road users and road users from other Member States. In Spain, Royal Decree 94/2006, of February 3, on the interoperability of electronic road toll systems in State roads.

On the other hand, in the proposal of multiple intersections in a network of urban roads (section 2.2 of this paper), as in the case of Madrid, for it to work it is essential that the three public administrations (State, Autonomous Community of Madrid or the municipality of Madrid) must coordinate, collaborate and cooperate (article 103 of the Spanish Constitution, article 3 of Law 40/2015, of October 1, the legal regime of the public sector). A single intersection manager of the Madrid road network should be created with the aim of improving traffic flow. However, in the proposal included in section 2.3 (fact 8) each intersection depends on a coordinator who makes decisions locally; that is, it does not coordinate with the other intersection managers. This situation goes against the principles of coordination, cooperation and collaboration between public administrations and violates the principle of efficiency in the allocation and use of public resources recognised in article 7 of the Organic Law 2/2012, of April 27, on Budgetary Stability and Financial Sustainability. Therefore, in a next work we will look for the formula of coordinating intersection networks to guarantee these principles and above all the traffic flow.

Therefore, the use of auction-based intersections does not violate freedom of movement and to circulate, principles of equality and non-discrimination. It even contemplates the possibility of crossing these intersections for free. EU law has shown that the establishment of tolls throughout the road infrastructure network does not violate this law when it is based on the "user pays" and "polluter pays" principles and the national legislation is not discriminatory (e.g., Directive 2011/76 /EU of the European Parliament and of the Council, of 27 September 2011, amending Directive 1999/62/CE on the charging of heavy goods vehicles for the use of certain infrastructures and Judgment of the Court of Justice of the European Union [Grand Chamber] June 18, 2019 Case C 591/17, Republic of Austria v. Federal Republic of Germany). The EU directives do not guarantee to EU citizens that, when exercising their right to free movement, it must be carried out free of charge. The economic policy of each EU member country must be able to determine whether such road infrastructure is financed by taxpayers or users. In short, it is an economic policy decision that corresponds to the Member States, therefore such road infrastructure can be financed by taxes or tolls (Opinion of Advocate General Wahl, delivered on 6 February 2019, Case C 591/17 Republic of Austria v. Federal Republic of Germany). In this way the tolls by auction add more price flexibility is adjusting supply and demand, establishing maximum and minimum prices.

---

[13] Judgment of the Spanish Supreme Court of February 15, 1996; Judgment of the Spanish Supreme Court of February 21, 1997, …

[14] Spanish Decree 215/1973, of 25 January, by which approves the content of the administrative specification sheets for the construction, conservation and operation of highways in concession contract.

[15] Its title is: "An Act to promote the safe integration of autonomous vehicles into the transportation system of the Commonwealth", 2017, Massachusetts. [https://malegislature.gov/Bills/190/S1945].

### 3.2. Human rights and legal principles related to auction-based intersections

The invention of the traffic light and its installation in the streets of the cities caused the modification of the laws, especially, of the traffic regulations. Until the regulation was not reformed, traffic lights could not regulate the circulation of vehicles. Intelligent road infrastructures, such as the intersection proposed in this paper for the circulation of fully autonomous vehicles, will imply a clear legal revolution of many regulations so that the installation and building of this auction-based intersection be allowed on roads and streets. The construction of these intelligent infrastructures will eliminate traffic lights and other traffic signals, and as numerous works have shown, these intelligent intersections reduce traffic congestion and prevent accidents (Dresner and Stone, 2008, pp 627-628; Beiker, 2012; Shubbak, 2013; Surden, 2016; Chen, Chen and Chen, 2017; Kockelman et al., 2018; among others). It would be necessary to unify the legal and technical solutions on the autonomous intersections from the international law as it happens with the traffic signals and the traffic lights that would come to replace. In the previous section we have outlined some regulatory changes so that these infrastructures can work, but also this type of infrastructures based on artificial intelligence in their source code must bear in mind a series of legal principles and values, as well as fundamental rights. For example, the European Parliament resolution of 16 February 2017, which includes recommendations to the Commission on civil law rules on robotics, suggests that research activities on robotics must respect fundamental rights. In this sense, the Spanish Constitution in article 18.4 provides: "The law shall limit the use of the computer science in order to guarantee the honour and personal and family privacy of citizens and the full exercise of their rights." In other words, the use of computer science cannot curtail the exercise of citizens' rights; constitutional limitation that must be very present in the regulation of artificial intelligence, also in these intelligent infrastructures. Bearing in mind this constitutional limitation, the auction-based intersections proposed in this paper must respect the right of free movement of citizens, as has been shown in section 3.1.

Liberty and Equality are closely linked. All citizens have the right to use public domain goods to guarantee freedom of movement. Provided there is no alternative route, the proposed algorithm must guarantee that the autonomous vehicle can cross the intersection free of charge. Fact that is not done currently (fact 6). And, in this regard, the proposed algorithm in this paper is not discriminatory and does not violate the article 14 of Spanish Constitution nor the anti-discrimination rules (Perez, 2016; Pearson, 2016; Levin, 2016; Council of Europe, 2018). The algorithm that these auction-based intersections use is neutral and refrain from the elaboration of user profiles;[16] that is, this algorithm does not correspond to what has been called "opaque algorithms" (Navas Navarro, 2017). In other words, we keep in mind in the design of these algorithms what is called "embedded values" (Surden, 2017). These values are widely used in the technological design choices made by engineers, for this reason the algorithm may end up having the effect of promoting or prioritizing certain social values over others or giving advantages or disadvantages to some social groups over others, especially in the process of auctioning the right and timing of passing through such autonomously managed intersections (for example, prioritizing the circulation of vehicles from some neighbourhood against others). Thus, for example, in the next regulation of autonomous vehicles of Massachusetts (An Act to promote the safe integration of autonomous vehicles into the transportation system of the Commonwealth, Bill S. no. 1945, 2017) it is said in the regulation of these vehicles that it will seek to protect the most affected and disadvantaged communities of the State and will ensure equal protection and the equitable distribution of the benefits and costs associated with the introduction of autonomous vehicles. In the proposal of this paper, the algorithm, in its initial formulation, is neutral because it does not incorporate promotional or discriminatory measures. We do not know what will happen when the algorithm starts to learn. (Caliskan, Bryson and Narayanan, 2017; Geslevich and Lev-Aretz, 2018). It is necessary to ensure that this algorithm is transparent, for example, through open code and guarantees that this algorithm can be audited.

---

[16] Regulation (EU) 2016/679 of the European Parliament and of the Council; of 27 April 2016; on the protection of natural persons with regard to the processing of personal data and on the free movement of such data, and repealing Directive 95/46/EC (General Data Protection Regulation) defines 'profiling' as "means any form of automated processing of personal data consisting of the use of personal data to evaluate certain personal aspects relating to a natural person, in particular to analyse or predict aspects concerning that natural person's performance at work, economic situation, health, personal preferences, interests, reliability, behaviour, location or movements."

When we translate a law (legal texts) to a formal rule (computer system) we can mask a series of subjective and debatable decisions about the meaning and scope of the formal rule (Pagallo and Durante, 2016), as we can choose between different legal options (section 3.1) in relation to issues such as public or private administration of the auction-based intersection manager, the free access during periods of time, the reversal of public domain goods or the crossing priority of some vehicles over others, etc. That is, the algorithms, being programmed by people, might have a strong ideological component that will have to be monitored and audited. The algorithm proposed in section 2 is subject to such monitoring and auditing. Finally, to avoid discrimination and algorithmic manipulation, it is necessary to audit the algorithms, when it includes the concepts of "algorithmic responsibility" and the "right to explanation (Kaminski, 2018); that is to say that there must be transparency and accessibility of the code (Goodman and Flaxman, 2017). The algorithm proposed in this paper is transparent and accessible. In summary, references to autonomous decision-making by artificial intelligence systems, including this autonomous infrastructure, cannot exempt the creators, owners and managers of these systems from responsibility for human rights violations, of the principle of non-discrimination and of other regulations, which could be committed with the use of these intelligent intersection systems (Etzioni, A. and Etzioni, O. (2016); Bench-Capon and Modgil (2017)). Consequently, this proposed algorithm must keep all information about the allocation of crossings, bids, etc. during a certain period of time (approximately 5 years) for facilitates the audit of the same and to verify that this algorithm responds to legality, it is accessible for a third party (human); it's transparent; recognise the right to judicial recourse to automated decisions (e-auctions) that do not violate human rights and do not discriminate for any reason.

As it has been revealed, the proposed algorithm must be designed and implemented to protect the personal data it uses, as required by Community law (Communication from the European Commission to the European Parliament, the Council, the European economic and social Committee and the Committee of the Regions, 2019). This proposed infrastructure would use databases, e.g. to store the already granted space/time reservations at the intersection managers, to keep track of the current reserve prices at the different intersections, etc. For this reason, its legal regulation is necessary in order to protect the privacy of data (article 10 of the Directive 2010/40/EU of the European Parliament and of the Council, of 7 July 2010, on the framework for the deployment of Intelligent Transport Systems in the field of road transport and for interfaces with other modes of transport). The Spanish Constitutional Court has indicated, in the Judgment no. 254/1993, that all citizens have the "right to control the use of the data inserted in a computer program". The constitutional case-law, and especially, the Judgment no. 292/2000, of November 30, has come to specify the scope, content and limits of this fundamental right to the protection of personal data. The Regulation (European Union) 2016/679 of the European Parliament and of the Council; of 27 April 2016; on the protection of natural persons with regard to the processing of personal data and on the free movement of such data, and repealing Directive 95/46/EC (General Data Protection Regulation) is the most important legal text in this matter. This European Regulation, which entered into force on May 25, 2018, indicates that necessary and proportionate measures must be contemplated to ensure the security of the network and of the information; that is, the ability of a network or information system to resist, at a certain level of trust, accidental events or illicit or malicious actions that compromise the availability, authenticity, integrity and confidentiality of personal data stored or transmitted, and the security of the services offered. This could include, for example, preventing unauthorised access to electronic communications networks and the malicious distribution of codes, and curbing "denial of service" attacks and damage to computer systems and electronic communications. The administrators of these auction-based intersection managers must guarantee the data security and protection, as well as having systems that avoid the hacking of the intelligent infrastructure (Taeihagh, Si Min Lim, 2019), as well as avoiding possible damages to the vehicles and to the people who use this infrastructure (auction-based intersection) with the objective that its use is completely safe[17]. It would be obligatory that intelligent infrastructures of this type have insurance, which entails that the insurance law regulates the clauses of this type of insurance contracts that cover the hypothetical economic liability derived from any failure of this intelligent intersection (Hernáez Esteban, 2018). There could be several assumptions of responsibility for the operation of the autonomous intersection (Lohmann (2016), Marchant and Lindor, 2012): a) If there is an accident due to failure of the coordinator or manager of intersections, for example, by giving simultaneous permission and two vehicles occupy the same space-time slot. The algorithm described in section 2 is designed so that this does not happen. The responsibility would correspond, in principle, to the intersection manager. b) The assumption that an accident occurs due to failure to comply with the conditions of crossing (space or time), the responsibility could be attributed to the constructor of the autonomous vehicle, c) if a breakdown occurs in the intersection, the intersection manager must avoid the accident and the responsibility could fall on the designer of the autonomous vehicle in accordance with the responsibility for damages caused by defective products. In any case, the algorithm proposed in section 2 must be implemented with the requirements of security, encryption and privacy required by the European Union regulations on data protection and cybersecurity.

This proposal of autonomously managed intersections based on auctions for reserving time and space must bear in mind that the outcome of these auctions would correspond to what has been called "smart contracts" (e-auctions) (Surden, 2012). Smart contracts have four characteristics: (i) have a computer program, (ii) be based on the blockchain technology (Millard, 2018; De Filippi and Wright, 2018; Górriz López, 2017), (iii) be self-executing and, (iv) be autonomous in the sense that human intervention is not necessary. The proposed algorithm can perfectly implement these four notes.

Differently from early proposals for autonomous intersection management in which the first vehicle that arrived was the first to cross the intersection (first in time, greater in right or First Come, First Served [Dresner and Stone, 2008; Vasirani and Ossowski, 2011, p 197; Parker and Nitschke, 2017]), in this proposal the intersection manager set prices for space/time reservations within a computational market that must be subject to legal principles of contractual regulations such as freedom of access to the reservation purchase system, publicity and transparency of electronic contracting procedures (public e-auctions), and non-discrimination and equal treatment between the buyers, among others. The algorithm proposed in section 2 must take into account all these legal requirements, avoiding discriminatory or dynamic prices according to user profiles (Zuiderveen Borgesius and Poort, 2017).

In the case proposed in this paper, it would be an automated contracting of both negotiating parties; that is to say, the contracting between intelligent agents (autonomous vehicle and auction-based intersection manager). The problems in this case are placed in relation to the risk distribution criteria in the event that there is an error in contracting or occurrence of fortuitous case as those described above (Urgern-Sternberg, 2018). The traditional category of vices of consent will have to be questioned because it is inadequate for a computational market whose actors are intelligent agents. This type of contracts between two intelligent agents raises many questions about the willingness of the parties, in accordance with the current contractual regulation. Therefore, a reform of contract laws would be compulsory (Navas Navarro, 2017).

## 4. Challenges for real-world-deployment

In this section we analyse certain important technical aspects that do not comply with the proposed system from the standpoint of Spanish legislation and in the context of international regulations, proposing solutions from the ethical and legal and/or technical scopes to guarantee its viability. We are aware that our proposed system is only a real approximation of as many others as possible in order to pave the way for the deployment in the real world of such systems in the near future.

As we have argued in the previous section, numerous legal problems need to be addressed in order to give legal support a smart traffic infrastructure including auction-based intersections. An exciting subject that exceeds the current regulation, where many sectors of the legal activity are intermingled: the own institution of the roads like goods of public domain and the use of the same one; competencies among public administrations; the public or private management of such infrastructures, freedoms and rights involved; the public contracts; vehicle passing preference and a long etcetera.

---

[17] Directive (EU) 2016/1148 of the European parliament and of the Council of 6 July 2016 concerning measures for a high common level of security of network and information systems across the Union and Commission Recommendation (EU) 2019/534 of 26 March 2019 Cybersecurity of 5G networks, C/2019/2335, *DO L 88 de 29.3.2019, p. 42/47.*

As we know, the intelligent infrastructure pursues intersection crossing assigning reserves to each vehicle. This operating system affects vehicles and traffic. Thus, from the point of view of the law, the regulation to cross the intersection in Spain must be carried out by the corresponding authority (state, autonomous or local Administrations), because there are different levels of competence.

In the sequel, we examine the different operational facts presented in section 2.3 and analyse their compliance with the current legal regulation. If needed, we suggest how the current law or technical solution should be adapted before such a system could be deployed in a real setting (Table 1 summarises this discussion):

- Fact 1: A vehicle can only pass through an intersection using space-time slots previously assigned by the intersection manager and using space quotas allocated at the specific time. From a legal point of view, this implies taking into account that roads are public domain goods, so implementations of intersection managers have to comply with traffic legislation. This is currently not the case for the system proposed in Section 2. In Spain, for instance, the Law on Traffic, Circulation of Motor Vehicles and Road Safety would need to be modified in order to allow the use of autonomous vehicles in their different modalities. In the realization of this reform would be intended to avoid violations that affect the freedom of movement and the principle of equality. It is a fact that the reserve system must be sustainable and not imply strong imbalances in terms of equality, so the indiscriminate use of reserves can be limited to a number of times per month or year. The purpose is to guarantee the access of the greatest number of users to the reservation system, provided they are autonomous vehicles (driver-less vehicle or with safety drivers).
- Fact 2: The intersection manager assigns a maximum of one vehicle to each space in a specific moment of time. This fact does not comply with the legislation, because the Spanish Law on Traffic, Circulation of Motor Vehicles and Road Safety does not include the possibility that there are managers of intersections. Again, in order to assure the legal viability of autonomous intersections in line with the proposal of Section 2, both norms would need to be modified.
- Fact 3: In order to claim a pass through an intersection, the vehicle must request space-time slots at a given time from the manager of the corresponding intersection. This fact does comply with the legislation so no modifications would be necessary.
- Fact 4: The allocation of space-time slots between vehicles is based on sealed-bid one shot auction. From the perspective of law, this constitutes an electronic auction regulated by the Spanish Law 33/2003 of November 3, on the Patrimony of Public Administrations and the Spanish Law of 9/2017 of November 8, Contracts of the Public Sector. Again, as to this aspect, the mechanism proposed in Section 2 would comply with Spanish legislation.
- Fact 5: In the proposed system, the intersection manager obtains benefits. From a legal point of view, the implementation of the intersection based on auctions, as described in Section 2, can be established in two ways: either through public management without benefits, or by means of private management with benefits through a concession. Both alternatives would comply with the current legislation in Spain.
- Fact 6: To participate in an auction, each intersection has a reserve price that has been published by the intersection manager. However, the offer of each vehicle must be higher than the price of the current reservation of the intersection. Here, the proposal of Section 2 does not comply with the legislation, roads are public domain goods and drivers' need to be given an option to use them free of charge. Here, we would favour a technical solution to comply with this legal requirement. The auction protocol could simply be modified such that drivers who have been waiting for a certain amount of time would be granted access to the intersection for free.
- Fact 7: The intersection managers introduced in Section 2 set reserve prices primarily to improve the distribution of traffic, rather than with the aim of obtaining benefits. Nonetheless, from the legal point of view, the price of the reservation is not limited and depends on the traffic demand. Still, the public Administration is competent to determine a maximum price and the algorithm introduced in section 2 should have such a limit. Still, this is not contemplated in the proposal outlined in Section 2, so it currently does not comply with Spanish legislation. An additional drawback pertains to the possibility of reserve price increasing continuously, if many vehicles wanted to use that intersection and there are no, or only very few, alternative routes. In this case drivers should be deterred from using the intersection. A higher price would certainty serve for this purpose but may not be socially and is definitely not legally acceptable. A solution to both aforementioned problems would consist in fixing a maximum reserve price for intersection and that, once this maximum price is reached at some intersection, it could be temporarily closed for drivers and would thus not accept new reservations for a certain amount of time.

- Fact 8: Each intersection manager makes decisions at the local level without explicit coordination with other managers. As outlined above, in cases with high traffic density reserve prices at all intersections in a certain area can grow. This behaviour does not comply with the legislation as, from the legal standpoint, the principles of efficiency of the infrastructure and cooperation between the different entities involved need to be respected. As above, setting a maximum price for intersection would be the easiest solution to put limits to such an "arms race" between different intersection managers. Furthermore, even though the mechanism outlined in Section 2 does not employ explicit communication between intersection managers, their reserve prices are coordinated implicitly through the environment.

Table 1. Summary of legal implications and possible solutions to the operational facts of the auction-based intersection system analysed in this paper.

| Fact | Description | Legal implications | Proposed solution |
|---|---|---|---|
| 1 | Vehicle can only cross an intersection using the space-time slots assigned by the intersection manager | Traffic law | Law modification: allow autonomous vehicles |
| 2 | The intersection manager assigns at most one vehicle to each space slot at each particular time | Law on Traffic, Circulation and Road Safety | Law modification: allow autonomous intersection coordinators |
| 3 | To cross an intersection, any vehicle has to request space-time slots from the intersection manager | Complies with current legislation | Not needed |
| 4 | Space-time slots assignment is based on a one-shot, sealed bid auction | Laws of Patrimony of Public Administrations and Contracts of the Public Sector | Not needed |
| 5 | The intersection manager obtains benefit | Public/private management | Not needed |
| 6 | Each vehicle's bid has to be higher than the current reserve price of the intersection | Free use of public goods | Technical: Free access to drivers who have been waiting certain time |
| 7 | The intersection manager aims at improving traffic distribution by modifying the reserve price. This is done locally, without maximum price | Public Administration is competent to set maximum price | Technical: fixing a maximum reserve price |
| 8 | Intersection managers do not explicitly coordinate with each other. Thus, with high traffic density, reserve prices at all intersections may grow | Efficiency of the infrastructure and cooperation between public administrations | Technical: fixing a maximum reserve price |

Next, we analyse five general assumptions that affect the proposed system and have an impact on the applicable law. The first assumption (assumption 1) determines that all vehicles that wish to cross an intersection must be able to communicate with the intersection manager and must have the software system installed. The regulation could, perfectly, include such an assumption, which would imply that vehicles cannot circulate without having the software system installed. However, socially this may not be an acceptable solution. From an ethical-legal standpoint and in order to avoid short-term inequalities, it would be better to establish a transition period where intersections can be used by both, vehicles that are able to use the proposed technology and those that are not. Technically, the proposed system could be adapted in this sense. The solution here consists in establishing two types of reservation time intervals that are alternated – one for the auction-based assignment and one for "traditional" vehicles. In fact, it would also be possible to change both intervals dynamically and adapt them to the current demand in each moment by monitoring the vehicles that want to pass through an intersection. Such monitoring could be done by combining communication with the vehicles with standard means of determining the number of arriving vehicles (e.g., through sensors). Dresner and Stone (2008) already analysed different techniques to allow for the coexistence of traditional and automatic vehicles in reservation-based infrastructures.

We have highlighted other general assumptions such as responsibility (assumption 2) and data protection (assumption 3). With regards to the former, in case of failure by the intersection manager (e.g., in the case that the manager has granted simultaneous permission to two vehicles to which a behaviour can be reported as compliance with some of the conditions of crossing) the fault will be yours when imposing civil and/or criminal liabilities. The assignment algorithm is designed such that it is not possible that two vehicles have the right to use the same space-time-slots, e.g., that two vehicles are authorised to cross the intersection at the same time. Another problem would be to regulate an alternative modus operandi in case an intersection manager is not working because of any failure. Such a case could be regulated in a similar way as it is done for standard traffic lights.

With regards to assumption 3, it is necessary to apply General Data Protection Regulation (EU) 2016/679 in order to improve cybersecurity and protection of data. The algorithms used within the computational mechanisms involving auction-based intersections must be effectively audited. The establishment of an Agency for Robotics and Artificial Intelligence, in Spain, would be desirable to this respect, following the recommendation of the European Parliament (Walker-Osborn, 2017; Kroll, Huey, Barocas, Felten, Reidenberg, Robinson, and Yu, 2017; Tutt, 2016). This Agency should dedicate itself to the supervision and audit of the algorithms (García Herrero, 2017) to, on the one hand, verify that the algorithms are a faithful translation of the regulations (including respect for human rights and the principle of non-discrimination) and, on the other, this algorithm audit must include the need to protect personal data. In addition, a certification mechanism must be used to prove compliance with these obligations and, consequently, a new work emerges: the algorithm inspector. In short, the algorithms must respond to the legality, the algorithms must be accessible so that a third party (human) can audit them, and the algorithms must be transparent. The right to apply to the courts must be recognised in the face of automated decisions that violate human rights, or are discriminatory or, in general, breach the legal system. Therefore, we suggest that the information of bids, assignments, etc. of each intersection manager be saved for a certain amount of time, which would need to be established at a normative level (usually 5 years).

Finally, we analyse two assumptions not included in the proposed system, but that must be taken into account from an ethical and legal point of view. In this context, the underlying concept of morality plays an important role as limit to the legal and ethical decisions of auction-based road intersections. According to Floridi and Sanders, an "agent is morally good if its actions all respect" the "threshold", that they call morality; "and it is morally evil if some action violates it. That view is particularly informative when the agent constitutes a software or digital system, and the observables are numerical" (2004, p 349). Here, we consider two especial cases. The first case (assumption 4), emergency vehicles (e.g. ambulances, fire trucks, police vehicles, etc.) that should have priority in the crossing, especially in emergency situations. In the second case (assumption 5), discounts/incentives can be applied for vehicles with special characteristics (e.g. low-emission cars, high-occupancy vehicle, vehicles by disabled drivers with different degrees of disability, etc.). Consequently, the legislator must establish promotion policies as a system of tax benefits for users who meet some of those characteristics.

Again, the work of Dresner and Stone (2008) already pointed to different technical option for implementing such priorities in an implementation of a reservation-based intersection. In general, the aforementioned two aspects could be solved in a common way by assigning priority levels to different types of vehicles (e.g., ambulances, police cars, etc. but also low-emission cars or cars owned by disabled). Such priority levels, for example, could establish discounts or weights on the reservation cost and the bidding price (in form of multipliers) or could even exempt certain vehicles from the task of bidding (e.g., for ambulances or other emergency vehicles). In fact, an assignment of mobility resources (like intersections), based on economic principles, is much more suitable for individualising the usage of such resources in order to meet certain global utility criteria.

# 5. Conclusions

The emerging field of AT aims at supporting large-scale open distributed systems, made up of smart software components that negotiate on behalf of their users. A key field of application are Smart Cities, and smart transportation in particular. Still, while the sandbox of AT provides models, methods, and tools to enable such applications from a technical point of view, several legal and ethical issues need to be addressed before the potential real-world deployment of such systems in a near future. In this paper, we used the approach of auction-based smart intersections as a case study to illustrate this matter. We argued that technically it is possible to support networks of such intersections, and to palliate the negative effect of providing preferential access to the infrastructure on social welfare through the dynamic adaptation of reserve prices. Then, we have analysed what would be the legal and ethical implications if such smart intersections were to be put into reality in the future. In particular, we have analysed what modifications would be needed in the smart traffic infrastructure itself, and what changes in legal regulations would be required before a potential real-world deployment could be considered.

The exercise we carried out is related but different to traditional legal requirements engineering where the objective is to determine the requirements a new system has to fulfil in order to comply with current regulations. Our objective, instead, was rather to determine the modification that should be introduced, both at the technical and the legal level, in order to put innovative solutions into practice. We think that this is a common problem and we have put forward some general steps that could help to address such problems.

Our future work will unfold among two major lines. Firstly, we will explore how the additional technical requirements can be integrated into the auction protocol for smart intersections, and re-evaluate the potential impact on their performance. In parallel, we will further look into the impact and consequences of the legal and ethical issues that have been identified. Secondly, we aim at performing studies similar to the one outlined in this article to other mechanisms, enabled by intelligent interactions with smart infrastructures, not only in the field of smart transportation, but also in domains such as smart grids and smart governance.

**Acknowledgments:** This work has been partially supported by the Spanish Ministry of Economy and Competitiveness, and the Spanish Ministry of Science, Innovation and Universities, co-funded by EU FEDER Funds, through grants TIN2015-65515-C4-X-R (*SURF*) and RTI2018-095390-B-C33 (InEDGEMobility).